\documentclass[11pt]{article}

\usepackage{cite}
\usepackage{amssymb}
\usepackage{amsmath}
\usepackage{bm}

\begin{document}

\title{Is the electron stationary in the ground state of the Dirac hydrogen atom in Bohm's Theory?}
\author{B. J. Hiley\footnote{E-mail address b.hiley@bbk.ac.uk.}.}
\date{TPRU, Birkbeck, University of London, Malet Street, \\London WC1E 7HX.\\ \vspace{0.4cm} }
\maketitle

\begin{abstract}
We show that, in the relativistic Bohm model of a Dirac-like particle, the electron in the ground state of the hydrogen atom is moving, unlike the prediction for the case of a Schr\"{o}dinger-like particle, where the electron is stationary.  This accounts for the empirically observed dilation of the decay time of the muon in the ground state of muonium.  

\end{abstract}

\section{Introduction}

The non-relativistic Bohm model of the ground state of the hydrogen atom finds the electron is not moving.  This seems a very odd result~\cite{ae53}.  Indeed for every excited $s$-state of the atom, the model predicts a stationary electron~\cite{gdmdgbh14}. The mathematical reason for this is that the wave function is real.  This means that in the polar decomposition of the wave function, $\psi(x,t)=R(x,t)\exp[iS(x,t)]$,  the phase,  $S$, is zero.  Thus when we use the so-called guidance condition, $p=\nabla S$, the momentum of the electron is zero.  Strange as this may be, it should be noted that the quantum expression for the current density  
\begin{eqnarray}
j=\frac{\hbar}{2im}[\psi^*(\nabla \psi)-(\nabla \psi^*)\psi]
\end{eqnarray} also vanishes so that standard quantum theory predicts that there is no probability current.  One could attempt to argue that this arises because the electron is executing a complex movement, but it is difficult to imagine how it could produce such a movement and produce zero angular momentum. 
To avoid such difficulties, Bohr argued that in quantum phenomena we should not attempt to give a picture of what the `electron is doing'.  All such `pictures' are ruled out because ``we are not dealing with an arbitrary renunciation of a more detailed analysis of atomic phenomena but with
a recognition that such an analysis is in principle excluded"~\cite{nb58}. 

In the Bohm model, following Mackey~\cite{gm63}, we assume that the local momentum, $p=\nabla S$, applies to each electron, so that we have a clear prediction.  Of course, we can ask for experimental evidence to decide which is the more correct picture, assuming we are allowed pictures in the standard approach.  Is there such an experiment?

If the electron is replaced by a muon, then we have an indirect way of telling if the muon is moving or not.  The muon is an unstable particle with a well-defined life time in its rest frame.   If it is moving fast enough, we will find a significant time dilation effect and the muon will live longer before it decays.  I have not been able to find a clean experiment that measures the life time of the muon in the ground state of a muon atom, however indirect evidence from muonium indicates that in fact the life time is longer \cite{muonlifetime}.  This result surely shows that the Bohm model must be wrong?

Before drawing such a conclusion we must realise that time dilation is a relativistic effect and so should we be reaching a conclusion that the model is wrong based on the use of the non-relativistic Schr\"{o}dinger equation?  Surely we should be using the Dirac equation. A first step towards a relativistic Bohm model was made by de Broglie \cite{dbr56} and discussed by Bohm and Hiley \cite{dbbh93} when they defined the velocity of the Dirac particle as 
\begin{eqnarray}
v^i=\frac{\bar\Psi \gamma^i\Psi}{|\Psi|^2}.  \label{eq:Diracv}
\end{eqnarray}

Recently Hiley and Callaghan~\cite{bhbc11, bhbc10} have shown in detail how the Bohm model can be formally extended to cover the Dirac equation.  There they also obtained expressions for the Bohm energy momentum current and quantum potential.  The detailed argument requires a  familiarity with orthogonal Clifford algebras to fully appreciate the extension, but fortunately we do not need to be familiar with all the details.  What we showed was  that the Bohm charge current is indeed given by equation (\ref{eq:Diracv}).   What we will show in the next section is that this current is not zero in the ground state of the muonium atom, so the relativistic Bohm electron will be moving with a finite velocity and this gives the required prediction of the dilated decay time of the muon.  

\section{The Ground State of the Hydrogen Atom}

\subsection{The Non-relativistic Schr\"{o}dinger Electron}

Let us quickly recall the Bohm approach to the hydrogen atom, and in particular the behaviour of the electron in the ground state.  We need first to solve the Schr\"{o}dinger equation
\begin{eqnarray*}
\frac{-\hbar^2}{2m}\nabla^2\psi(r,\theta, \phi)-\frac{e^2}{r}\psi(r,\theta, \phi)=E\psi(r,\theta, \phi).
\end{eqnarray*}
The energy eigenfunctions for the hydrogen atom~\cite{lb90} are then
\begin{eqnarray*}
\psi_{nlm}(r,\theta,\phi)=-\left[\frac{4(n-l-1)!}{(na_0)^3n[(n+1)!]^3}\right]^{1/2}\rho^lL^{2l+1}_{n+1}(\rho)e^{-\rho/2}Y^m_l(\theta, \phi)
\end{eqnarray*}
where $\rho=2r/na_0$, and $a_0=\hbar^2/me^2$.  Fortunately we simply need the ground state which can be directly read off the general equation so that
\begin{eqnarray*}
\psi_{100}=(\pi a_0^3)^{-1/2}e^{-r/a_0}.
\end{eqnarray*}
This wave function is real with the phase $S=0$.  Thus the Bohm momentum of the ground state electron is $p_B=\nabla S=0$.  Thus the electron is at rest. Indeed this feature is the same in all $s$-states.  As soon as one goes to the $l\ne0\ne m$, the electron is then in motion.  One can explore the details of this motion from examining each state separately.  We will not analyse this behaviour here.

It is the stationary electron in the $s$-states that causes an uneasy feeling as one's classical intuition demands the electron should move in a planetary-like orbit.  But, as we have already remarked, even then there is a conflict with our classical intuition of angular momentum.  In the $s$-state the electron has no angular momentum and should not be rotating around the nucleus.  However  in muonium,  the muon in the ground state must be moving because its half-life is dilated.  This is a relativistic effect, so let us now turn to the Dirac equation.

\subsection{The Relativistic Dirac Electron}

In order to work out the Dirac current we need an expression for the relativistic ground state of the hydrogen atom.  This has been worked out many times and we will write down the expression given by Bjorken and Drell \cite{jbsd64} for hydrogen-like atoms.  With the Coulomb potential written in the form $V(r)=-\frac{\alpha}{r}$ the ground state for spin up is given by
\begin{eqnarray*}
\Psi_{n=1,j=1/2,\uparrow}(r,\theta, \phi)=A(r)  \begin{pmatrix} 
      1 \\
      0 \\
      i\frac{(1-\gamma)}{\alpha}\cos\theta\\
       i\frac{(1-\gamma)}{\alpha}\sin\theta e^{i\phi}
   \end{pmatrix}.
\end{eqnarray*}
The ground state for spin down is given by
\begin{eqnarray*}
\Psi_{n=1,j=1/2,\downarrow}(r,\theta, \phi)=A(r)   \begin{pmatrix} 
      1 \\
      0 \\
      i\frac{(1-\gamma)}{\alpha}\sin\theta e^{i\phi} \\
      - i\frac{(1-\gamma)}{\alpha}\cos\theta 
   \end{pmatrix},
\end{eqnarray*}
where $A(r)=\frac{(2m\alpha)^{3/2}}{\sqrt{4\pi}}\sqrt{\frac{1+\gamma}{2\Gamma(1+2\gamma)}}\mbox{\small$(2mZ\alpha r)^{\gamma-1}$}e^{-m\alpha r} $.\newline
[Our units are $\hbar=c=1$,\;\;$\alpha = e^2/4\pi,\;\;\gamma=\sqrt{1-\alpha^2}$].

For simplicity we will write the spin up solution in the  simpler form
\begin{eqnarray*}
\Psi_{n=1,j=1/2,\uparrow}(r,\theta, \phi)=A(r)   \begin{pmatrix} 
      1 \\
      0 \\
      iB\\
      iDe^{i\phi}
   \end{pmatrix},
\end{eqnarray*}
where we have written
\begin{eqnarray*}
B=\frac{(1-\gamma)}{\alpha}\cos\theta\quad \mbox{and}\quad D=\frac{(1-\gamma)}{\alpha}\sin\theta.
\end{eqnarray*}
We then write 
\begin{eqnarray*}
\bar\Psi=\Psi^\dag\gamma^0= \begin{pmatrix}       1, & 0, & -iB, & -iDe^{-i\phi} \\
      \end{pmatrix}   \begin{pmatrix} 
                  1 & 0 & 0 & 0 \\
            0 & 1 & 0 & 0 \\
            0 & 0 & -1 & 0\\
            0 & 0 & 0 & -1\\
         \end{pmatrix},
\end{eqnarray*}
so that the current can be evaluated using 
\begin{eqnarray*}
j^\mu=A^2(r)\begin{pmatrix}       1, & 0, & -iB, & -iDe^{-i\phi} \\
      \end{pmatrix}\gamma^\mu \begin{pmatrix} 
      1 \\
      0 \\
      iB\\
      iDe^{i\phi}
   \end{pmatrix}.
\end{eqnarray*}

After some tedious work, we find for the ground state with spin up,
\begin{eqnarray*}
j^1=-2A^2(r)D\sin\phi;\quad j^2=2A^2(r)D\cos\phi;\quad j^3=0.
\end{eqnarray*}
While for the ground state with spin down.
\begin{eqnarray*}
j^1=2A^2(r)D\sin\phi;\quad j^2=-2A^2(r)D\cos\phi;\quad j^3=0
\end{eqnarray*}
with \begin{eqnarray*}
D=\left(\frac{1-\gamma}{\alpha}\right)\sin\theta.
\end{eqnarray*}
If we fix $\theta$ and project the motion onto the $x-y$ plane, we find the spin up state rotates anti-clockwise, while the spin down state rotates clockwise.

Since the charge current is simply related to the velocity of the electron by $v^i=j^i/|\Psi|^2$, we immediately see that the relativistic electron is moving in the relativistic Bohm model.  It  clearly rotates either anti-clockwise if its spin is up,  or clockwise if it is in the spin down state.  Thus in the case of muonium, we see that the muon is also moving and therefore subject to a time dilation.  This then explains why we see the muon live longer in the atomic state. Since we are essentially using the standard formalism to reach our conclusions, the actual value of the dilation will be the same as predicted using the standard approach.  

It is interesting to note that in the non-relativistic limit
\begin{eqnarray*}
\gamma\rightarrow 1;\quad \frac{1-\gamma}{\alpha}\rightarrow 0;\quad D\rightarrow 0; \quad v\rightarrow 0.
\end{eqnarray*}
Thus in the non-relativistic limit, when the Schr\"{o}dinger equation holds, the electron is stationary.

\section{Conclusion}

Thus we see that there is no inconsistency between the relativistic Bohm model and the experimental evidence on the decay time of the muon in muonium.  Time dilation is a relativistic effect and it therefore  must be discussed in a relativistic context.  The details of the relativistic treatment of the Dirac electron can be found in  Hiley and Callaghan \cite{bhbc11, bhbc10}.



\bibliography{myfile}{}
\bibliographystyle{plain}

\end{document}